\documentclass[letters,fleqn,usenatbib]{mnras}

\usepackage[T1]{fontenc}
\usepackage{ae,aecompl}

\usepackage{graphicx}	% Including figure files
\usepackage{amsmath}	% Advanced maths commands
\usepackage{amssymb}	% Extra maths symbols

\newcommand{\planck}{\textit{Planck}}
\newcommand{\al}{Airlander 10}
\newcommand{\alcmb}{AirlanderCMB}

\newcommand{\bicep}{{\sc BICEP1 }}

\newcommand{\piper}{PIPER }

\newcommand{\class}{CLASS }

\newcommand{\sigr}{\ensuremath{\sigma_{r=0}}}
\newcommand{\sigtau}{\ensuremath{\sigma_\tau}}
\newcommand{\lmin}{\ensuremath{\ell_{\rm min}}}

\newcommand{\forecast}{{\tt CMB4CAST}}
\newcommand{\camb}{{\tt CAMB}}
\newcommand{\am}{{\tt am}}

\title[CMB at Cruising Altitude]{Cosmic Microwave Background Science at Commercial Airline Altitudes}

\author[S. M. Feeney et al.]{Stephen M. Feeney,$^{1,2}$
Jon E. Gudmundsson,$^{3,4}$
Hiranya V. Peiris,$^{3,5}$\thanks{E-mail: h.peiris@ucl.ac.uk (HVP)}
\newauthor
Licia Verde,$^{6,7}$
and Josquin Errard$^{8,9}$
\\
$^{1}$Department of Physics, Blackett Laboratory, Imperial College London, Prince Consort Road, London SW7 2AZ, UK\\
$^{2}$Center for Computational Astrophysics, 160 5th Avenue, New York, NY 10010, USA\\
$^{3}$The Oskar Klein Centre, Department of Physics, AlbaNova, Stockholm University, SE-106 91 Stockholm, Sweden\\
$^{4}$Nordita (Nordic Institute for Theoretical Physics), Roslagstullsbacken 23, SE-106 91 Stockholm, Sweden\\
$^{5}$Department of Physics and Astronomy, University College London, Gower Street, London WC1E 6BT, UK\\
$^{6}$ICC, University of Barcelona, IEEC-UB, Mart\`i i Franqu\'es, 1, E08028 Barcelona, Spain\\
$^{7}$ICREA, Pg. Llus Companys 23, 08010 Barcelona, Spain\\
$^{8}$Sorbonne Universit\'es, Institut Lagrange de Paris (ILP), 98 bis Boulevard Arago, 75014 Paris, France\\
$^{9}$LPNHE, CNRS-IN2P3 and Universit\'es Paris 6 \& 7, 4 place Jussieu, F-75252 Paris Cedex 05, France
}

\date{Accepted XXX. Received YYY; in original form ZZZ}

\pubyear{2016}

\begin{document}
\label{firstpage}
\pagerange{\pageref{firstpage}--\pageref{lastpage}}
\maketitle

% Abstract of the paper
\begin{abstract}
Obtaining high-sensitivity measurements of degree-scale cosmic microwave background (CMB) polarization is the most direct path to detecting primordial gravitational waves. Robustly recovering any primordial signal from the dominant foreground emission will require high-fidelity observations at multiple frequencies, with excellent control of systematics. We explore the potential for a new platform for CMB observations, the \al\ hybrid air vehicle, to perform this task. We show that the \al\ platform, operating at commercial airline altitudes, is well-suited to mapping frequencies above $220$ GHz, which are critical for cleaning CMB maps of dust emission. Optimizing the distribution of detectors across frequencies, we forecast the ability of \al\ to clean foregrounds of varying complexity as a function of altitude, demonstrating its complementarity with both existing (\planck) and ongoing (C-BASS) foreground observations. This novel platform could play a key role in defining our ultimate view of the polarized microwave sky.
\end{abstract}

% Select between one and six entries from the list of approved keywords.
% Don't make up new ones.

%%%%%%%%%%%%%%%%%%%%%%%%%%%%%%%%%%%%%%%%%%%%%%%%%%

%%%%%%%%%%%%%%%%% BODY OF PAPER %%%%%%%%%%%%%%%%%%

\section{Introduction}

Measurements of the large-scale polarization of the cosmic microwave background (CMB) are central to two key goals of cosmology: detecting primordial gravitational waves and robustly measuring the summed neutrino masses. The limiting factor in achieving these aims, obscuration by polarized Galactic emission, can be controlled using the differing frequency dependence of the Galactic and cosmological components. Obtaining high-sensitivity, multi-frequency CMB polarization measurements on large scales is, however, a serious experimental challenge.

CMB polarization can be decomposed into $E$ modes, sourced predominantly by scalar perturbations, and $B$ modes, generated at arc-minute scales by gravitational lensing~\citep{Zaldarriaga1998} and potentially at degree scales by gravitational waves excited during inflation~\citep{Seljak1997,Kamionkowski1997}. Though the inflationary signal could be undetectably small (or non-existent), its singular nature makes it an irresistible target. The primordial signal is potentially detectable at $\ell \lesssim 200$ before it becomes subdominant to lensing. Large-scale $E$ modes provide a direct handle on the optical depth to reionization, $\tau$, degeneracies with which fundamentally limit our ability to measure neutrino masses with the CMB~\citep{Smith2006}. 

Polarized foregrounds---dominated by  synchrotron and dust emission---overwhelm large-scale primordial $B$ modes. These foregrounds can be reasonably captured with models described in~\citet{PlanckX2015}, although weak evidence is beginning to emerge for greater complexity, with spatially-varying frequency dependence~\citep{PlanckL2016} or multiple dust components~\citep{Meisner2015,Draine1998}. Ground-based observations of polarized dust emission are made difficult by the  opacity of the Earth's atmosphere above $220$ GHz. This  can be overcome by deploying a payload on a stratospheric balloon or satellite.

The types of CMB observatories currently available---ground-based observatories, stratospheric balloons, and satellites---offer distinct pros and cons. Ground-based observatories allow the shortest construction timescales, the use of cutting-edge technology, regular upgrades and repairs. 
Telescope diameter is relatively unrestricted compared with stratospheric and satellite instruments, making the ground the location of choice for observing the smallest angular scales. 
These observatories are, however, fixed at a particular location within Earth's atmosphere, restricting the accessible portion of the sky and constraining observations to atmospheric frequency windows centred on roughly 40, 100, 150, and 250~GHz \citep{Hanany2013}. Even within these windows, measurements on the largest scales are subject to location-specific systematic effects such as ground pickup, wind, and turbulence, preventing ground-based experiments from probing  the largest angular scales. 

By contrast, satellites can probe the entire CMB electromagnetic spectrum free from atmospheric contamination, but they typically undergo 10--20~year development cycles. The prospect of near-space-like conditions has pushed many scientific collaborations to employ stratospheric balloons, which typically operate at altitudes of 30--40~km. Their  reduced cost  results in shorter development cycles and  exploitation of more modern technology. Unfortunately, typical stratospheric balloon flights, so-called long-duration balloons (LDB), have a mean flight time of 20~days. Further,  because payloads sustain non-negligible damage during landing, experiments normally do not fly more than once every two years, yielding $\sim$10~days of integration per year. Upcoming experiments hope to exploit newly developed superpressure-balloons (SPB) which are expected to provide $\sim100$-day stratospheric flights with somewhat reduced lift capacity \citep{Eberspeaker2011}.

Another sub-orbital platform, helium airships operating at commercial jet altitudes, might emerge  as a novel platform for CMB observations. In this {\it Letter}, we explore the possibility of mapping the high-frequency microwave sky at degree scales from one such airship: the \al\ hybrid air vehicle. 

%%%%%%%%%%%%%%%%%%%%%%%%%%%%%%%%%%%%%%%%%%%%%%%%%%

\section{Airlander 10}

\begin{figure}
\center 
\includegraphics[width=\columnwidth]{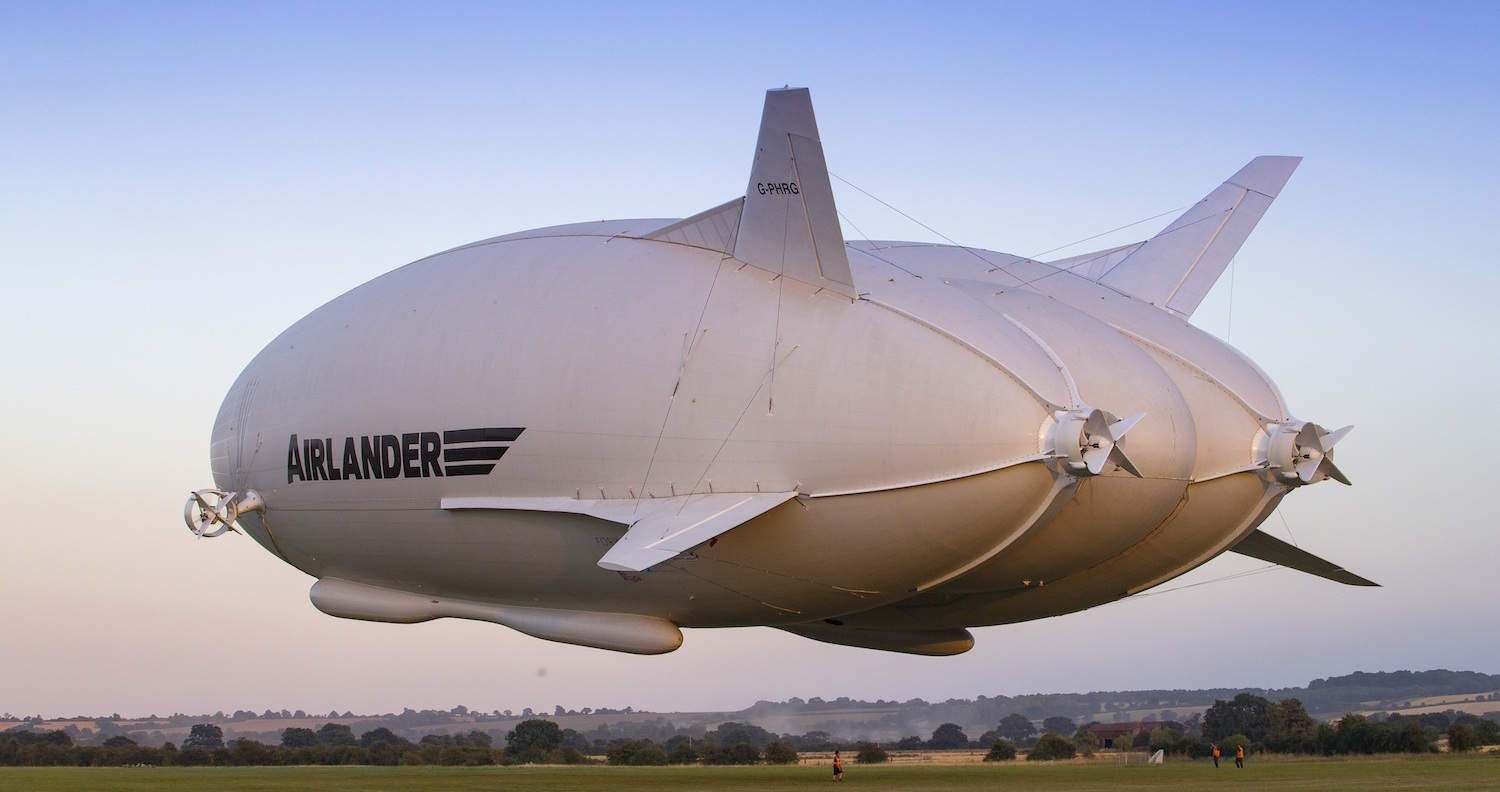}
\caption{\al's first flight on August 17th, 2016.}
\label{fig:airlander}
\end{figure}

\al, developed by the United Kingdom-based company Hybrid Air Vehicles (HAV), is the largest aircraft currently flying (see Fig.~\ref{fig:airlander}).  \al\ generates 60\% of its lift through helium buoyancy and the remainder through its aerodynamic shape. Its four engines, with vectored thrust, allow the aircraft to maintain position at a minimum airspeed of 20 knots. Designed to meet the heavy lifting and long-duration surveillance and communications needs of the military and commercial aviation sectors, \al's basic scope is to heavy lift cargo of up to 10,000 kg at altitudes of up to 10,000 feet (3 km) or to carry a maximum surveillance payload of 1,000 kg at altitudes of up to 20,000 feet (6 km) for up to three weeks (via remote piloting), without the need for an airport. Its maximum flight altitude could potentially be increased to 30,000 ft (9 km) with further modifications and testing, and careful operation from a suitable 8--10,000 feet operating base.

There are several features of \al\ that lend it well to CMB observations. \al's potential payload mass and flight duration are both comparable to LDB, but the turnaround time for \al\ flights could be days rather than years, with minimal risk of payload damage during landing. As a result, \al\ has the potential to carry LDB-esque CMB observatories, offsetting the reduction in altitude with increased integration time. HAV indicate that \al's basic design could be modified to mount a CMB telescope on the aircraft's lower fins, yielding an instantaneous field of view of around a quarter of the sky. Should civil regulations for remote piloting of \al\ be in place by 2019, HAV anticipate that a remotely piloted flight would be feasible within five years. We assume such a scenario is plausible without attempting to assess the technical challenges associated with deploying a telescope on this platform. From here onwards, we refer to this modified aircraft as \alcmb. \alcmb's utility in obtaining large-scale polarization information rests on the frequency range attainable at commercial airline altitudes: here we investigate this in detail.

%%%%%%%%%%%%%%%%%%%%%%%%%%%%%%%%%%%%%%%%%%%%%%%%%%

\section{Forecast Methodology}

Our methodology is based on the \forecast\ code~\citep{ErrardFeeney2016} which forecasts the ability of a given experimental configuration to isolate lensing and primordial $B$ modes in the presence of Galactic foregrounds and hence constrain cosmology. Foreground cleaning is assumed to be performed by a parametric maximum-likelihood approach~\citep{Brandt1994,Eriksen2006,Stompor2009}, in which the frequency dependence of each foreground component (which may vary spatially) is estimated from multi-frequency observations and then used to construct a cleaned CMB map. The noise and foreground residuals in the resulting CMB map~\citep{Errard2011,Errard2012} are propagated through the rest of the forecast~\citep{Verde2006}, and projected constraints on cosmological parameters are obtained with a power-spectrum-based Fisher formalism.

Our forecasts depend on two fundamental components: a sky model and an instrument model. For the former, we assume the polarized microwave sky takes the simplest form supported by current data: the CMB obscured by Galactic synchrotron and dust emission. \forecast\ requires spatial templates for each sky component, along with a parametric form for their frequency and angular-scale dependence. We source spatial templates from~\citet{PlanckX2015,PlanckIX2015}, define the dust and synchrotron frequency dependences as a modified grey-body (temperature $T_{\rm d} = 19.6$ K and spectral index $\beta_{\rm d} = 1.59$) and a power law (spectral index $\beta_{\rm s} = -3.1$), respectively, and model their multipole dependence as $C_\ell^{\rm d} \propto \ell^{-2.4}$ and $C_\ell^{\rm s} \propto \ell^{-2.6}$~\citep{PlanckIntXXII2015,PlanckIntXXX2016}. Spectral indices are assumed constant over $15^\circ$ patches. The CMB power spectra required by \forecast\  are computed using \camb~\citep{Lewis2000}, assuming the best-fit cosmology from {\it Planck}'s ``TT+lowP+lensing+ext'' analysis~\citep{PlanckXIII2015}.

To fully specify an instrument model, \forecast\ requires the observable fraction of sky ($f_{\rm sky}$), the range of accessible multipoles and the central frequency, bandpass, angular resolution and white noise level of each channel. The white noise level of an array of $N_{\rm det}$ bolometers is given by $N_\ell = { 4 \pi f_{\rm sky} s^2 }( N_{\rm det} t_{\rm obs} )^{-1}$, where $t_{\rm obs}$ is the  observation time and $s$ is the polarized detector sensitivity. 

{\bf Detector Modelling: } We adopt a relatively simple model to predict detector sensitivities as a function of altitude and therefore atmospheric loading. This model assumes that the detectors are cooled to 100~mK to minimize detector noise due to stochastic thermal fluctuations (phonon noise) and adopts optical properties similar to those deployed by the \bicep experiment \citep{Takahashi2010} (see Table~\ref{tab:detector_props}). The general detector performance assumed for our fiducial instrument can be extrapolated from the estimated properties of current and future instruments, including \piper and \class \citep{Gandilo2016,Hileman2014}

\begin{table}
	\centering
	\caption{Detector properties assumed for this analysis. Loading corresponds to the assumed total optical loading from instrument, before accounting for optical efficiencies.}
	\label{tab:detector_props}
	\begin{tabular}{ccccccccc}
		\hline
		Band center & Bandwidth  & FWHM & Opt. eff & Loading\\
		{[GHz]} & [GHz] & [arcmin] & & [pW] \\
		\hline
		40 & 10 & 90 & 0.30 & 5.7 \\
		94 & 23 & 43 & 0.30 & 5.7  \\
		150 & 36 & 30 & 0.30 & 10.0 \\
		220 & 66 & 21 & 0.30 & 16.7 \\
		270 & 81 & 15 & 0.30 & 15.7 \\
		350 & 28 & 14 & 0.30 & 7.3 \\		
		\hline
	\end{tabular}
\end{table}

\begin{figure}
\includegraphics[width=\columnwidth]{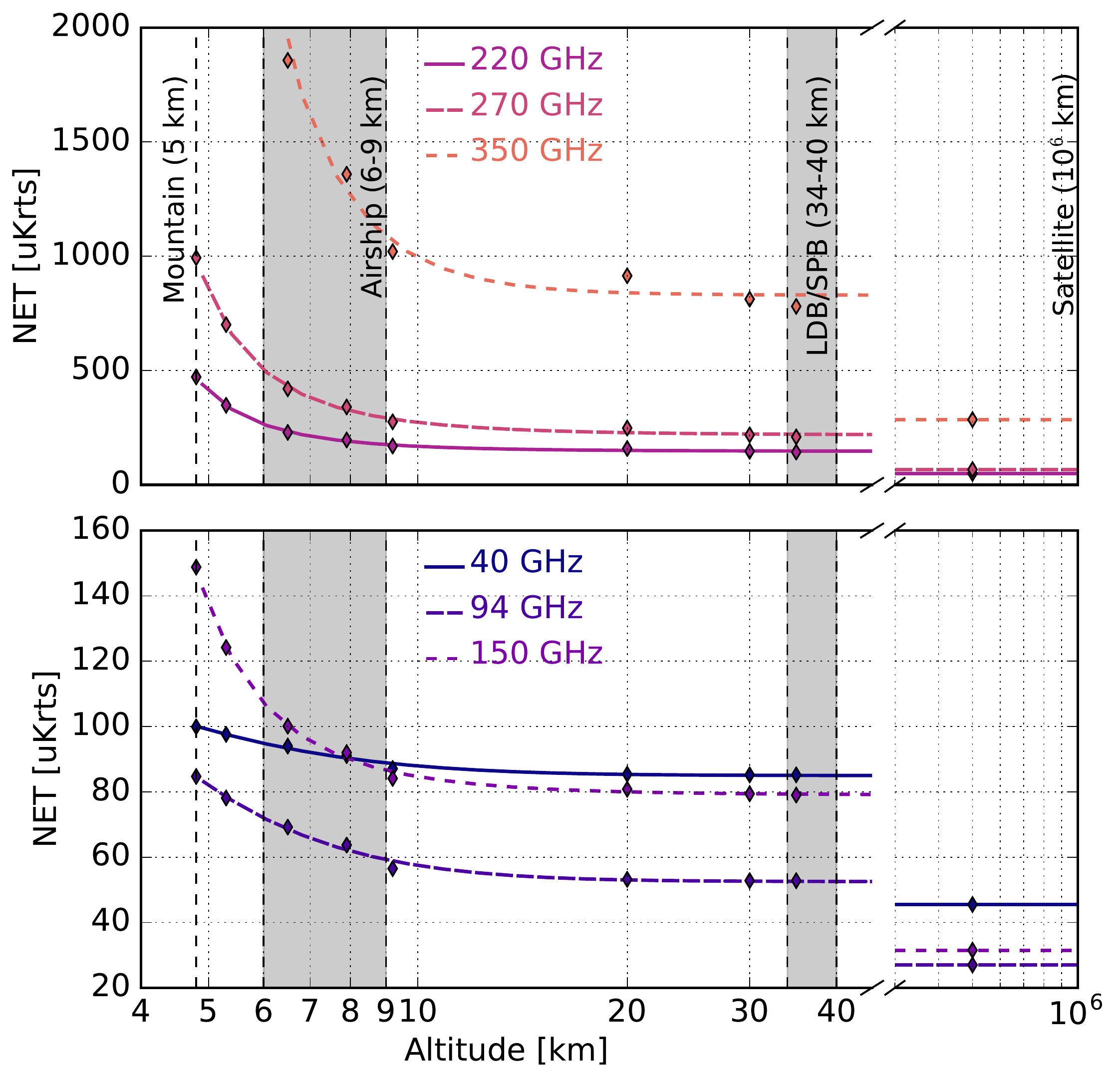}
\caption{Simulated sensitivity of a single detector in each frequency channel as a function of altitude above the Atacama plateau (mountain). Solid lines are power-law fits to the discrete model predictions, plotted as diamonds. Estimates are produced assuming 1~mm PWV and a viewing angle of 45$^\circ$.}
\label{fig:sens_vs_alt}
\end{figure}

We refer the reader to \citet{Mather1982,Lamarre1986,IrwinTES} for details of detector sensitivity modeling. For the six frequencies considered, noise contributions from thermal fluctuations and readout electronics are assumed to be constant: 3.8 and 6.0~aWs$^{1/2}$, respectively. Conversion from W to ${\rm K_{CMB}}$ is performed using the properties in Table \ref{tab:detector_props} assuming the detectors are single-moded. Using the \am\ Atmospheric Model package \citep{Paine2016} to produce an estimate of the atmospheric spectral radiance as a function of altitude, we vary detector loading due to atmosphere while fixing photon loading contributions from instrument emission and the CMB. Atmospheric loading is derived assuming 1~mm of precipitable water vapor (PWV) and 45$^\circ$ elevation angle.  Since \am\ provides location-specific results, we use model predictions pertaining to the atmosphere above the Atacama; however, results should not differ significantly for other locations if we ignore PWV variability. Note that 1~mm PWV is almost certainly greater than the median PWV value across any given year at the site of the Atacama Cosmology Telescope \citep{Dunner2013}. Fig.~\ref{fig:sens_vs_alt} shows the predicted detector sensitivity as a function of altitude for this model.

Our detector sensitivity model is generated assuming the sensitivity of suborbital instruments will continue to be limited by their optical components and not the statistical properties of incoming CMB photons. The predictions are based on a few basic assumptions about instrument performance: these assumptions can be contested, and should not be interpreted as statements of fact.

{\bf Experimental Setup:} We assume a basic experimental setup of 10,000 detectors in total, comparable to proposed LDB missions such as EBEX-IDS and BFORE~\citep{Hanany2015,Niemack2016}. We consider six frequency bands, centred on 40, 94, 150, 220, 270, and 350 GHz, with Gaussian beams whose widths (FWHMs) are reported in Table~\ref{tab:detector_props}. Bands with $\nu \le 220$ GHz can be observed from the ground without penalty (Fig.~\ref{fig:sens_vs_alt}); indeed, AdvACTPol, BICEP-Keck, Simons Array and SPT-3G plan to. Here, we model all channels flying at altitude to avoid having to select a particular observatory to optimize against. We assume all experiments considered can scan half the sky and access multipoles in the range $20 \le \ell \le 2500$ unless otherwise stated. We do not explicitly treat systematic effects, which are highly experiment-specific; rather, we assume such effects will render unusable the modes on scales larger than the minimum multipole considered. The total mission observation time is taken to be 12 weeks with an observing efficiency of 80\%. To make best use of existing data, we include \planck\ in all \forecast\ forecasts as in \citet{ErrardFeeney2016}.

%%%%%%%%%%%%%%%%%%%%%%%%%%%%%%%%%%%%%%%%%%%%%%%%%%

\section{Analysis \& Results}

Fig.~\ref{fig:sens_vs_alt} shows that the sensitivity of the 40, 94, and 150 GHz channels only changes by a factor of $\sim$2 from the highest ground-based observatories to the stratosphere; however, the sensitivity improves swiftly with altitude for frequencies above 220 GHz. \alcmb\ becomes a competitive platform at frequencies of 220 GHz and above.

Next, we investigate the ability of complete experiments to clean foregrounds and constrain cosmology. We begin with a toy experiment with 10,000 detectors equally distributed across the six frequency channels, prior to optimization. As the instrumental resolution is low and the post-component-separation CMB map is noise- rather than lensing-limited, we do not consider delensing in these forecasts. We constrain the simplest extension of the standard cosmological model containing inflationary $B$ modes and take the marginalized uncertainty on the tensor-to-scalar ratio ($r$), $\sigr$, as our performance metric.
\begin{figure}
\includegraphics[width=\columnwidth]{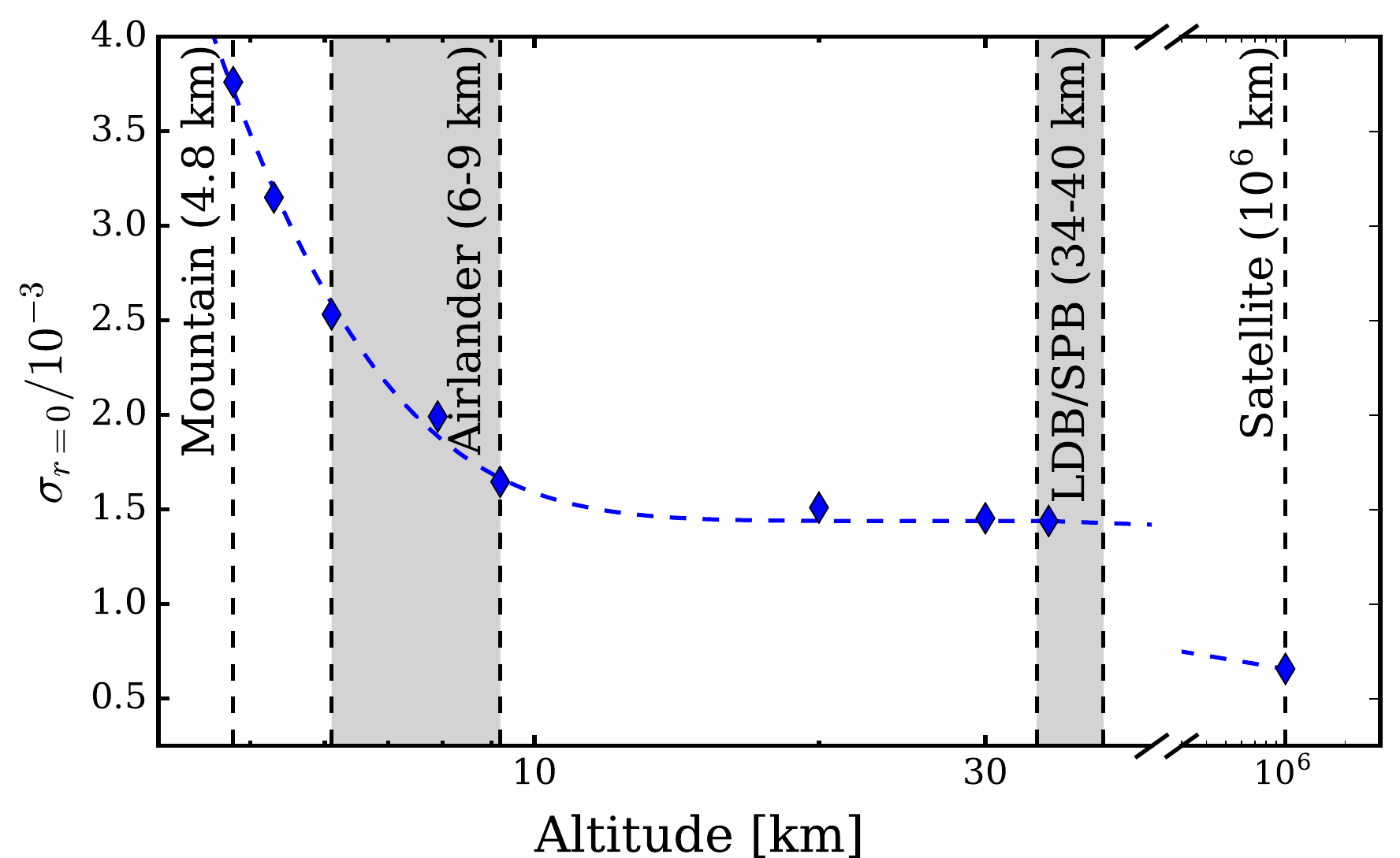}
\caption{Constraints on the tensor-to-scalar ratio for an unoptimized 10,000-detector toy experiment operating at various altitudes. Synchrotron and dust contamination have been removed with the aid of \planck\ data; observation time is taken to be 12 weeks at 80\% efficiency.}
\label{fig:sig_r_vs_alt}
\end{figure}
We plot the performance of our toy experiment as a function of altitude in Fig.~\ref{fig:sig_r_vs_alt}. The impact of the additional frequency coverage is clear, as the constraints tighten by a factor of 2.3 flying at maximum altitude compared to Atacama: over 60\% of the improvement of space.

\begin{figure*}
\includegraphics[width=2\columnwidth]{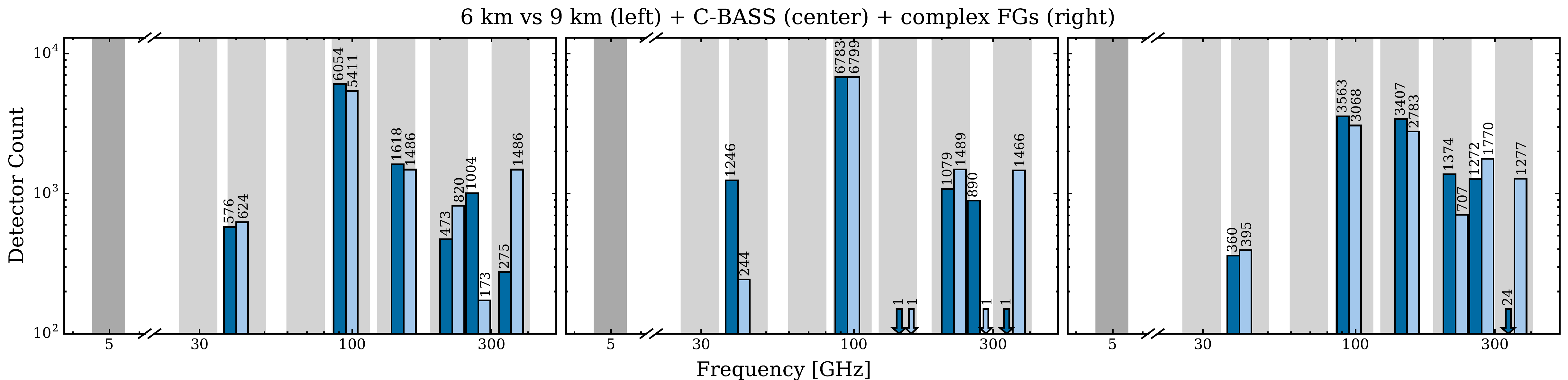}
\caption{Optimal detector distributions for \alcmb\ operating at 6 and 9 km (dark and light blue) in various settings: cleaning simple foregrounds in combination with \planck\ (left) and both \planck\ and C-BASS (centre); cleaning complex foregrounds in combination with \planck\ and C-BASS (right). \planck\ and C-BASS channels are displayed as light- and dark-grey bands, respectively.}
\label{fig:optimized_configurations}
\end{figure*}

We now optimize the frequency coverage as a function of altitude, following~\citet{Errard2011}. Using these optimal detector distributions, we further investigate the ability of \alcmb\ to clean our fiducial foregrounds and constrain cosmology as a function of its minimum multipole, \lmin. Our baseline is that all modes with $\ell \ge 20$ can be observed reliably; we also consider scenarios in which \lmin\ is 3 or 50. The optimal distributions are plotted in Fig.~\ref{fig:optimized_configurations} (left), with corresponding parameter constraints reported in Table~\ref{tab:configs_and_constraints}. Focusing initially on the detector distributions, we note that the differences between altitudes are largely confined to the highest frequencies. In both cases the bulk of detectors are placed in the most sensitive channel, 94 GHz, and the detectors at frequencies up to 220 GHz are distributed roughly as 1:10:2.5:1. Beyond 220 GHz, however, the impact of altitude is apparent: the 350 GHz channel is the main dust monitor at 9 km, but not 6 km.

\begin{table}
	\centering
	\caption{Parameter constraints for \alcmb\ as a function of altitude, \lmin, additional data, and foregrounds. We consider fiducial (fid.), flat and complex (comp.) foregrounds; the model assumed during optimization (opt. FGs) may be incorrect.}
	\label{tab:configs_and_constraints}
	\begin{tabular}{ccccccc}
		\hline
		Alt. &  $\lmin$ & Inc. & Opt. & True & \sigr & \sigtau \\
		{[km]} & & C-BASS & FGs & FGs & $[10^{-3}]$ & $[10^{-3}]$ \\
		\hline
		6 & 20 & & fid. & fid. & 1.64 & 4.21 \\
		 & 3 & & fid. & fid. & 1.59 & 2.66 \\
		 & 50 & & fid. & fid. & 1.95 & 4.21 \\
		 & 20 & \checkmark & fid. & fid. & 1.16 & 4.21 \\
		 & 20 & \checkmark & fid. & flat & 1.35 & 4.41 \\
		 & 20 & \checkmark & fid. & comp. & 32.9 & 72.5 \\
		 & 20 & \checkmark & comp. & comp. & 21.8 & 64.6 \\
		 & 20 & \checkmark & comp. & fid. & 1.41 & 4.21 \\
		\hline
		9 & 20 & & fid. & fid. & 1.14 & 4.20 \\
		 & 3 & & fid. & fid. & 1.14 & 2.62 \\
		 & 50 & & fid. & fid. & 1.40 & 4.20 \\
		 & 20 & \checkmark & fid. & fid. & 0.856 & 4.18 \\
		 & 20 & \checkmark & fid. & flat & 0.947 & 4.39 \\
		 & 20 & \checkmark & fid. & comp. & 33.5 & 72.0 \\
		 & 20 & \checkmark & comp. & comp. & 16.6 & 59.1 \\
		 & 20 & \checkmark & comp. & fid. & 1.02 & 4.18 \\
		\hline
	\end{tabular}
\end{table}

Turning to the parameter constraints, we find that \alcmb\ can place one-sigma limits of $\sigr = 1.6\times10^{-3}$ when operating at 6 km, or $1.1\times10^{-3}$ at 9 km. The limits do not change greatly if \lmin\ is reduced to 3---the experiment is foreground-limited on large scales---but degrade by 20\% if modes with $\ell<50$ are unusable. Constraints on $\tau$, which are sourced by large-scale $E$-mode measurements, are insensitive to \alcmb's altitude but depend strongly on \lmin. If \alcmb\ can access modes with $\ell \ge 3$ it will constrain $\tau$ with an uncertainty of $2.7\times10^{-3}$; if \lmin\ is beyond the reionization bump, the uncertainty grows to $4.2\times10^{-3}$.

Since \alcmb's strength is its high-frequency observations, we investigate its complementarity with an experiment offering superb low-frequency data: C-BASS~\citep{King2010}. C-BASS is gathering full-sky observations at 5 GHz~\citep{Irfan2015} with a 45~arcmin beam and 4500 $\mu$K-arcmin noise; here we assume it can cover the \alcmb\ patch down to \lmin\ = 20. The results of combining with C-BASS and \planck\ are reported in Fig.~\ref{fig:optimized_configurations} (centre) and Table~\ref{tab:configs_and_constraints}. For both altitudes, the main effect is to concentrate detectors into fewer bands: the 150 GHz channel becomes obsolete, its detectors redistributed to the 94 and 220 GHz bands. Since the 150 GHz band is less sensitive to the CMB than 94 GHz, it is disfavoured when C-BASS is present to regulate residuals. There are again interesting differences between baseline and maximum altitudes. At 6 km, the number of 40 GHz detectors increases, which aids in marginalizing over the foreground residuals; the noisy 350 GHz channel is dropped. At 9 km, however, adding C-BASS prioritizes 94 GHz over 40 GHz (reducing the noise of the main CMB channel) and moves all detectors from 270 GHz to 220 GHz (allowing better control of foreground residuals).

There is considerable uncertainty in current foreground models, so we investigate the impact of optimizing using incorrect foregrounds. First, we optimize using the correct model but incorrect parameter values. We run a single forecast (without reoptimizing) in which the foreground spectral indices are flattened (within their one-sigma \planck\ limits) to be closest to the CMB ($\beta_{\rm d}=1.55$, $\beta_{\rm s}=-2.7$). In this case, constraints on $r$ degrade by only 10--16\%. Next, we consider the case in which the model itself is incorrect, and the spectral indices vary rapidly, requiring estimation in each pixel. In this case, the failure is catastrophic: constraints degrade by a factor of 30--40. If we reoptimize assuming the complex foreground model we find minima in which all channels are populated~(Fig.~\ref{fig:optimized_configurations}, right), though constraints on $r$ are still heavily degraded, with \sigr\ in the range 1.7--2.2$\times10^{-2}$. However, these configurations are still able to place constraints on the {\em simple} foreground model of 1.0--1.4$\times10^{-3}$: only a 20\% penalty. Operating at altitude is essential in this case: dropping the three highest frequency channels penalizes constraints by 63\%.

%%%%%%%%%%%%%%%%%%%%%%%%%%%%%%%%%%%%%%%%%%%%%%%%%

\section{Conclusions}

With limited information about the nature of polarized foregrounds, the size, location and optimal frequency coverage of the ideal sky regions for $B$-mode observations remain largely unknown. To some extent, current and future experiments must gamble on these critical design parameters~\citep{Kovetz2015,Kovetz2016}. With relatively fast turnaround times, airships like \alcmb\ offer a modular platform that can be quickly optimized for varied sky regions and frequency ranges.

Optimizing \alcmb's frequency coverage for a range of altitudes, foreground scenarios and complementary datasets yields two important conclusions. First, realization of \alcmb's maximum altitude is not critical to the success of the platform, though, if possible, constraints on $r$ improve by 35\%. Second, optimizing for the most pessimistic foreground scenario does not significantly penalize performance if foregrounds are more benign. By contrast, constraints degrade by a factor of over 30 if the foregrounds are incorrectly assumed to be simple. As bands below 270 GHz will be observed from the ground, \alcmb\ need carry only 1300-3000 detectors to realize this performance.

High-altitude observations represent a critical step in preparing the ground for a space mission tasked with delivering the ultimate CMB polarization data. In addition to demonstrating detector technology, such observations will characterize polarized foregrounds in the detail required to ensure the design of the space mission suits the science requirements. Given the cost and development time of a space mission, novel platforms such as \alcmb\ could play a very important role in defining our ultimate view of the polarized microwave sky.

%%%%%%%%%%%%%%%%%%%%%%%%%%%%%%%%%%%%%%%%%%%%%%%%%

\section*{Acknowledgements}

We thank Stevie Bergman, Jeffrey Filippini, William Jones,  Lyman Page, and  George Efstathiou for useful discussions. SMF was partially supported by STFC in the United Kingdom. The Center for Computational Astrophysics is supported by the Simons Foundation. HVP was partially supported by the European Research Council under the European Community's Seventh Framework Programme (FP7/2007-2013) ERC GA n. 306478-CosmicDawn. LV acknowledges support by  Spanish MINECO under projects AYA2014-58747-P and MDM-2014-0369 of ICCUB (Unidad de Excelencia Maria de Maeztu). JEG acknowledges support by Katherine Freese through a grant from the Swedish Research Council (Contract No. 638-2013-8993). The work of JE was performed within the Labex ILP (reference ANR-10-LABX-63) part of the Idex SUPER, and was supported by the ANR, as part of the programme Investissements d'avenir under the reference ANR-11-IDEX-0004-02.

%%%%%%%%%%%%%%%%%%%%%%%%%%%%%%%%%%%%%%%%%%%%%%%%%%
%%%%%%%%%%%%%%%%%%%% REFERENCES %%%%%%%%%%%%%%%%%%
\bibliographystyle{mnras}
\bibliography{airlander_cmb}
%%%%%%%%%%%%%%%%%%%%%%%%%%%%%%%%%%%%%%%%%%%%%%%%%%
% Don't change these lines
\bsp	% typesetting comment
\label{lastpage}
\end{document}